\documentclass[a4paper,11pt]{article}
\pdfoutput=1 % if your are submitting a pdflatex (i.e. if you have
\usepackage{jheppub} % for details on the use of the package, please
\usepackage[T1]{fontenc} % if needed

\newcommand\be{\begin{equation}}
\newcommand\ba{\begin{eqnarray}}
\newcommand\ee{\end{equation}}
\newcommand\ea{\end{eqnarray}}

\newcommand\lp{\left(}
\newcommand\rp{\right)}
\newcommand\lb{\left[ }
\newcommand\rb{\right] }
\newcommand\lv{\left\{ }
\newcommand\rv{\right\}}

\newcommand{\nn}{\nonumber}

\title{\boldmath Thermal Loop Effects on Large-Scale Curvature Perturbation in the Higgs Inflation}

\author[a,b,c]{Po-Wen Chang,}
\author[c,d]{Cheng-Wei Chiang,} 
\author[d,e]{and Kin-Wang Ng}

\affiliation[a]{Center for Cosmology and AstroParticle Physics (CCAPP), The Ohio State University,\\Columbus, Ohio 43210, USA}

\affiliation[b]{Department of Physics, The Ohio State University,\\Columbus, Ohio 43210, USA}

\affiliation[c]{Department of Physics, National Taiwan University,\\Taipei 10617, Taiwan}

\affiliation[d]{Institute of Physics, Academia Sinica,\\Taipei 11529, Taiwan}

\affiliation[e]{Institute of Astronomy and Astrophysics, Academia Sinica,\\Taipei 11529, Taiwan}

\emailAdd{chang.1750@osu.edu}
\emailAdd{chengwei@phys.ntu.edu.tw}
\emailAdd{nkw@phys.sinica.edu.tw}

\abstract{It is known that the Higgs potential in the Standard Model can drive a successful inflation as long as the Higgs field couples non-minimally to gravity. It is then inevitable to take into account the loop corrections of the Standard Model particles to the Higgs potential in the Higgs inflation. In this paper, we discuss the one-loop corrections at finite temperature to the curvature perturbation generated during the Higgs inflation. We find that the thermal loop effects can suppress the power of the curvature perturbation at large scales, thus resulting in a low quadrupole of the temperature anisotropy in the cosmic microwave background.}


\begin{document} 
\maketitle
\flushbottom

\section{Introduction}\label{1_intro}

Over the past few decades, the cosmic inflation has become a well-accepted solution to the horizon, flatness and monopole problem in cosmology \cite{Starobinsky1980,Guth1981,Linde1981,Linde2008}. The simplest theoretical picture of inflation consists of a scalar field $\phi$ ({\it i.e.}, the inflaton) rolling slowly over a flat potential $V(\phi)$, which could mimic a nearly constant vacuum energy and produce an exponential expansion of the Universe. It has been found that the quantum fluctuations during inflation typically result in a nearly scale-invariant power spectrum of the curvature perturbation. The prediction generally fits the observational data of the temperature fluctuations in the cosmic microwave background (CMB) radiation very well \cite{Boyle2006}, making the inflation even more robust to occur in the early Universe. 

Despite successfully predicting many observational features, the flat potentials in most inflationary models are not natural in particle physics. Without understanding the fundamental principles leading to a flat potential, the theory of inflation is merely a phenomenological description of the early Universe. Among the inflationary models on the market, the Higgs inflation scenario treats the Higgs field $h$ as the inflaton and provides an elegant interpretation of the origin of the flat potential by virtue of a quadratic non-minimal coupling between $h$ and gravity \cite{Bezrukov2007,Bezrukov2013}. After performing a conformal transformation, it is found that the Higgs potential becomes asymptotically flat in the large-field regime and thus can be treated by the standard procedures in the slow-roll approximation. It turns out that the Higgs inflation is favored by the current Planck constraints as long as the non-minimal coupling constant $\xi$ is as large as $\sim 10^{4}$. More detailed analyses and discussions of the Higgs inflation can be found in refs.~\cite{Bezrukov2011,George2014,Allison2014,Hertzberg2010,Burgess2010}.

From the theoretical point of view, the Higgs inflation is appealing, as it is well-motivated by the Standard Model (SM) of particle physics. It not only relates the fundamental physics at microscopic scales to the cosmological observations at the largest scales, but also facilitates a possibility to constrain the mass of the top quark with cosmological observables \cite{Bezrukov2009,Bezrukov2009-2,Salvio2013}. A recent study also finds that, by taking the renormalization group (RG) running of the SM coupling constants into account, an inflection point may exist in the Higgs potential, thus allowing an ultra-slow-roll phase during which high peaks in the curvature power spectrum can be generated. This provides a realization of the origin of primordial black holes that may account for the binary black hole merger events in the Advanced LIGO/Virgo observations~\cite{Ezquiaga2018}.

In most inflationary scenarios, effects of the thermal bath would be omitted. It is because the embedding physics of the inflation models is scarcely known or it is assumed that any non-zero temperature prior to inflation would drop exponentially once inflation begins. However, we shall show that the temperature effect can be particularly important to the Higgs inflation. According to finite-temperature field theory~\cite{Dolan1974,Carrington1992,Quiros1999,Laine2016}, the Higgs field would acquire quantum loop corrections to its free energy density due to a non-zero temperature, leading to a temperature-dependent effective potential. Such a thermal effect is crucial to inducing the electroweak phase transition (EWPT). It is commonly believed that below a critical temperature $T_{c}\simeq 150 ~\mathrm{GeV}$, the Higgs field bears a spontaneous symmetry breaking $SU(2)_{L} \times U(1)_{Y} \rightarrow U(1)_{\mathrm{EM}}$ and subsequently finds its non-zero vacuum expectation value $v$ \cite{KolbTurner}. Motivated by the theory of the EWPT associated with the Higgs field, it is natural to consider the Higgs field in thermal equilibrium with a heat bath of non-zero temperature before the inception of the Higgs inflation ({\it i.e.}, in a pre-inflationary thermal bath) and scrutinize how the finite-temperature effective potential can leave an imprint on the cosmological observables. 

Previous studies such as refs.~\cite{Bhattacharya2006,Powell2007,Wang2008,Das2015} have shown that the mode function of the inflaton 
quantum perturbation during inflation can be non-trivially modified if there is a pre-inflationary radiation-dominated epoch at non-zero temperature. The modified mode function can lead to the suppression of the resulting perturbation power spectrum on large angular scales. In ref.~\cite{Das2015}, the authors have further considered the effect of the thermal initial state, which indeed enhances the power on large angular scales; nevertheless, they have shown that the non-trivial mode dynamics effectively overrides the effect of the thermal initial state. In this work, we firstly take into account the thermal effects on the inflation by calculating the thermal loop effective potential of the Higgs field in the conformal (Einstein) frame, following the well-established formalism in the thermal field theory and the Bunch-Davies vacuum. Here we choose the Bunch-Davies vacuum rather than the thermal vacuum in order to manifest the one-loop thermal corrections. We find that the Higgs inflation with non-zero temperature tends to suppress the amplitude of the primordial curvature power spectrum $P_{\mathcal{R}}(k)$ at large scales. Since the temperature decreases exponentially soon after the outset of inflation, the suppression in the power spectrum would also decrease. Consequently, the theory predicts a lower $P_{\mathcal{R}}(k)$ for small $k$ modes that smoothly returns back to the zero-temperature Higgs inflation. It has long been observed that there is a lack of power in the quadrupole moment of the CMB angular power spectrum~\cite{Oliveira-Costa2004}, dubbed the quadrupole anomaly. In this study, we shall show that the thermal loop effect on the Higgs inflation can serve to explain the quadrupole anomaly of the CMB radiation.\footnote{Some theoretical scenarios have also been proposed to understand the low quadrupole.  See, for example, refs.~\cite{Contaldi2003,Lopez2019} and the references therein.}

This paper is organized as follows: we first give an essential overview of the Higgs inflation in section~\ref{2}, and then develop an approach to performing the one-loop thermal correction to the Higgs inflation in section~\ref{3}. After analyzing the general properties of the one-loop effective potential at finite temperature in the Einstein frame, we calculate $P_{\mathcal{R}}(k)$ and the TT (temperature) angular power spectrum $\mathcal{D}^{TT}_{\ell}$ in section \ref{4}. We show that our results fit better to the temperature angular power spectrum extracted from the Planck $2018$ data than the best-fit base-$\Lambda$CDM cosmology due to the suppression of power at large scales. In section \ref{5}, we summarize the conclusions of this work. We provide useful relations of the physical quantities and equations between the Jordan frame and the Einstein frame in appendix \ref{A}.  Throughout this paper, we adopt $(-+++)$ as the metric sign convention and the natural unit $\hbar = c = k_{B} = 1$. We also use the Friedmann-Lema\^{i}tre-Robertson-Walker (FLRW) metric $ds^2 = g_{\mu\nu} dx^{\mu}dx^{\nu} = -dt^2 + a(t)^2 d\vec{x}^2 $.

%=================================================================%
\section{Higgs inflation}\label{2}
%=================================================================%

In the SM, the tree-level potential of the Higgs field $h$ is\footnote{The Higgs field $h$ here represents the radial mode of the SM Higgs doublet: $\mathcal{H}=(0,~h)^\mathrm{T}/\sqrt{2}$.}
\be  \label{JVh}
V_{0}(h)=\frac{\lambda}{4}(h^2-v^2)^2~,
\ee
where the vacuum expectation value is estimated to be $v \simeq 246~\mathrm{GeV}$. A na\"{i}ve attempt to take the Higgs field as an inflaton with the quartic potential (\ref{JVh}) would generally fail, as the density perturbation generated by the inflaton require the Higgs self-coupling $\lambda \sim 10^{-13}$ to be consistent with the CMB observations, whereas it is about $10^{-1}$ as inferred from the measured Higgs bosons mass $m_h = 125 ~\mathrm{GeV}$ \cite{CMS2012,ATLAS2012,PDG2018}. The Higgs inflation \cite{Bezrukov2007,Bezrukov2013}, on the other hand, remediates the problem by coupling the scalar Higgs field to the spacetime geometry. In this model, the action associated with the Higgs field is 
\be \label{SJh}
S_{J}= 
\int d^4x \sqrt{-g} ~\left[\frac{M^2}{2}f(h)R -\frac{1}{2}g^{\,\mu\nu}\partial_{\mu}h\,\partial_{\nu}h -V_{0}(h)    \right]~,
\ee
where the quadratic form of the non-minimal coupling is\footnote{The quadratic non-minimal coupling can be realized by the quantum field theory in curved spacetime \cite{Birrell1982,Buchbinder1992,Muta1991}. The application of the general theory with non-minimal coupling ({\it i.e.}, the scalar-tensor theory) in cosmology can be found in ref.~\cite{Faraoni2004}.}
\be  \label{fh}
f(h)=1+\frac{\xi}{M^2}h^2~,
\ee
with a coupling constant $\xi$ and a mass parameter $M$ defined phenomenologically by the reduced Planck mass through $M^2_{P}=1/(8\pi G) \equiv M^2+\xi v^2$. As we will see in the following, $\xi v^2$ is always much smaller than $M_{P}^2$ for the values of $\xi$ that we are interested in. Consequently, we will directly replace $M$ with $M_{P}$ hereafter.

The action in eq.~\eqref{SJh} describes the Higgs field in the Jordan frame (JF) --- the frame with an explicit non-minimal coupling term $f(h)R$. The field equations in the JF is relatively complicated. To facilitate the analyses of inflationary dynamics, it is convenient to recast the action in the Einstein frame (EF) via a conformal transformation
\be \label{CF0}
g_{\mu\nu} ~~~~\rightarrow~~~~ \widetilde{g}_{\mu\nu}=f(h)\,g_{\mu\nu}~.
\ee
We can then rewrite $S_{J}$ in a canonical form\footnote{The tilde denotes the quantities defined in the EF. For the relations of various physical quantities between the Jordan and the Einstein frames, see appendix~\ref{A}.}
\be  \label{SEh}
\int d^4x\sqrt{-\widetilde{g}} ~\lb  \frac{M_{P}^2}{2}\widetilde{R}-\frac{1}{2}\widetilde{g}^{\,\mu\nu}\partial_{\mu}\chi\,\partial_{\nu}\chi -U_{0}(\chi)\rb
~,
\ee
with the EF scalar field $\chi$ defined by
\be  \label{EFchi}
\frac{d\chi}{dh}= \lp\frac{f+3M^2_{P}\cdot f'^{\,2}/2}{f^2}\rp^{1/2} = \lb \frac{1+(1+6\,\xi)\,\xi\, h^2/M^2_{P}}{\lp 1+\xi h^2/M^2_{P}\rp^2}\rb^{1/2}
\ee
and the EF potential at tree level given by
\be  \label{EFU}
U_{0}\lp h(\chi)\rp =f^{-2}V_{0}(h) =\frac{\lambda}{4}\cdot \lb \frac{ h(\chi)^2-v^2}{ 1+\xi h(\chi)^2/M^2_{P}~}\rb^2 ~.
\ee
For $\xi\gg 1$, the parameter range of interest to us, eq.~(\ref{EFchi}) can be easily solved in the large-field limit $h\gg M_{P}/\xi$ as:
\be\label{chiapprox}
\frac{d\chi}{dh} \approx \displaystyle \frac{\sqrt{6}\,\xi h / M_{P}}{1+\xi h^2/M^2_{P}} ~~~ \Rightarrow ~~~ \chi(h) \approx \displaystyle \sqrt{\frac{3}{2}}M_{P} \,\mathrm{ln} f(h) ~.
\ee
It is also useful to set up the following identity from eq.~(\ref{chiapprox}):
\be
f\lp h(\chi)\rp \approx \mathrm{exp}\lp \frac{2\,\chi}{\sqrt{6}M_{P}} \rp~.
\ee
Therefore, in the large-field regime ($h\gg M_{P}/\xi$) we have
\be  \label{hlarge}
h(\chi) \approx \frac{M_{P}}{\sqrt{\xi}}\,\lb \mathrm{exp}\lp \frac{2\,\chi}{\sqrt{6}M_{P}} \rp-1 \rb^{1/2}~,
\ee
and the EF potential in eq.~(\ref{EFU}) becomes
\be \label{Ularge}
U_{0}(\chi) \approx \frac{\lambda M^4_{P}}{4\,\xi^2}\,\lb 1-\mathrm{exp}\lp \frac{-2\,\chi}{\sqrt{6}M_{P}} \rp \rb^2~.
\ee
Note that here the vacuum expectation value of the Higgs field $v$ has been ignored because we are considering $h \gg M_{P}/\xi \gg v$. The potential in eq.~(\ref{Ularge}) plays a critical role in validating the Higgs inflation. When $h \gg M_{P}/\sqrt{\xi}$ or $\chi \gg \sqrt{6}M_{P}/2$, the potential becomes asymptotically flat and approaches a constant $\lambda M^4_{P}/4\,\xi^2$. It is straightforward to estimate the potential slow-roll parameters for $h \gg M_{P}/\sqrt{\xi}$ in the EF by eqs.~(\ref{EFchi}) and (\ref{EFU}), and the standard slow-roll approximation demands:
\ba
\epsilon_{\mathrm{v}} &&= \frac{M^2_{P}}{2}\lp\frac{U_{0}'}{U_{0}}\rp^2 \approx \frac{4M^4_{P}}{3\,\xi^2\,h^4} \ll 1~;
\label{epsilonHiggs}
\\
\left| \eta_{\mathrm{v}} \right| &&= \left| M^2_{P}\,\frac{U_{0}''}{U_{0}} \right|\approx \frac{4M^2_{P}}{3\,\xi\,h^2} \ll 1~,
\label{etaHiggs}
\ea
where each prime denotes a derivative with respect to $\chi$ henceforth. Eqs.~(\ref{epsilonHiggs}) and (\ref{etaHiggs}) provide the sufficient conditions for successful inflation to occur. In general, the inflation ends when $\epsilon_{\mathrm{v}}(h_{\mathrm{end}}) = \epsilon_{\mathrm{v}}(\chi_{\mathrm{end}}) \simeq 1$.

In order to account for the horizon and the flatness problems, the inflation should last at least $\sim 57.7$ \emph{e}-folds after the horizon exits the CMB scale ($k_{\mathrm{CMB}}\sim 2 \times 10^{-4} ~h\mathrm{Mpc}^{-1}$). Defining the EF conformal time $d\tau = \sqrt{f(h)}\,dt $ and $\tau_{\mathrm{CMB}}$ as the time of CMB horizon crossing in the EF, it is easy to obtain $\chi_{\mathrm{CMB}}$ by estimating the \emph{e}-folds from $\tau_{\mathrm{CMB}}$ to the end of inflation via the EF potential
\ba
\widetilde{N}\lp\tau_\mathrm{CMB},\,\tau_{\mathrm{end}}\rp &&= \int^{\tau_{\mathrm{end}}}_{\tau_{\mathrm{CMB}}} \widetilde{H}\,d\tau \approx \frac{1}{M_{P}^2}\int^{\chi_{\mathrm{CMB}}}_{\chi_{\mathrm{end}}} \frac{U_{0}}{U_{0}'}\,d\chi ~,   \label{NEformJ}
\ea
where the definition of $\widetilde{H}$ is given in eq.~(\ref{AppH}). Demanding $\widetilde{N} \simeq 57.7 $ and applying the approximation in eq.~(\ref{Ularge}) to eq.~(\ref{NEformJ}), we get
\ba
57.7 &&\simeq \frac{\sqrt{6}}{4 M_{P}} \int^{\chi_{\mathrm{CMB}}}_{\chi_{\mathrm{end}}} \lb \, 1-\mathrm{exp}\lp\frac{-2\,\chi}{\sqrt{6}M_{P}}\rp \, \rb \mathrm{exp}\lp\frac{2\,\chi}{\sqrt{6}M_{P}}\rp \,d\chi   \nn
\\
&& = \left.  \frac{\sqrt{6}}{4M_{P}} \lb ~\frac{\sqrt{6}M_{P}}{2}\,\mathrm{exp}\lp \frac{2\,\chi}{\sqrt{6}M_{P}} \rp  -\chi ~ \rb      \right|^{\chi_{\mathrm{CMB}}}_{\chi_{\mathrm{end}}}  ~. \label{NEformE}
\ea
The field value $\chi_{\mathrm{end}}$ at the end of inflation is determined by
\be
\epsilon_{\mathrm{v}}(\chi_{\mathrm{end}}) = \left. \frac{M_{P}^2}{2} \lp \frac{U_{0}'}{U_{0}} \rp^2  \right|_{\chi_{\mathrm{end}}} 
\approx \frac{4}{3}\,\mathrm{exp}\lp \frac{-4\,\chi_{\mathrm{end}}}{\sqrt{6}M_{P}} \rp  \lb\, 1 -\,\mathrm{exp}\lp \frac{-2\,\chi_{\mathrm{end}}}{\sqrt{6}M_{P}} \rp \,  \rb^{-2} \simeq 1 ~.  \label{chiend}
\ee
Eqs.~(\ref{NEformE}) and (\ref{chiend}) together give $\chi_{\mathrm{CMB}} \simeq 5.41~M_{P}$ and $\chi_{\mathrm{end}} \simeq 0.94~M_{P}$.\footnote{The field values $\chi_{\mathrm{CMB}}$ and $\chi_{\mathrm{end}}$ here numerically correspond to $h_{\mathrm{CMB}} \simeq 9.14 ~M_{P}/\sqrt{\xi} $ and $h_{\mathrm{end}} \simeq 1.04 ~M_{P}/\sqrt{\xi} $, respectively. We have checked that eq.~(\ref{Ularge}) remains a good approximation at $h \simeq M_{P}/\sqrt{\xi} $ as long as $\xi \gg 1$. In contrast to the Higgs field $h$ in the JF, the predictions of $\chi_{\mathrm{CMB}}$ and $\chi_{\mathrm{end}}$ from eqs.~(\ref{NEformE}) and (\ref{chiend}) are independent of the value of $\xi$ that one chooses.} Following the procedure in ref.~\cite{Bezrukov2007}, we can use the WMAP normalization \cite{WMAP} to fix the relation between $\xi$ and $\lambda$
\be
\frac{U_{0}(\chi_{\mathrm{CMB}})}{\epsilon_{\mathrm{v}}(\chi_{\mathrm{CMB}})} \simeq (0.0274~M_{P})^4  ~.  \label{WMAPNorm}
\ee
Making use of eqs.~(\ref{epsilonHiggs}) and (\ref{WMAPNorm}) and $\lambda = m_{h}^2/2v^2 \simeq 10^{-1}$ \cite{CMS2012,ATLAS2012,PDG2018}, we get $\xi \approx 4.7 \times 10^4 \sqrt{\lambda} \sim 10^4$, a large non-minimal coupling. The spectral index $n_{s}$ and the tensor-to-scalar ratio of perturbation $r$ can also be estimated in a straightforward manner. See ref.~\cite{Bezrukov2013} for further discussions of the Higgs inflation.

%=================================================================%
\section{Thermal effects on the Higgs potential \label{3}}
%=================================================================%

%==============
\subsection{The one-loop thermal correction: Choices of the frames \label{31}}
%==============

\begin{figure}[ht]
  \centering
  \includegraphics[width=0.8\textwidth]{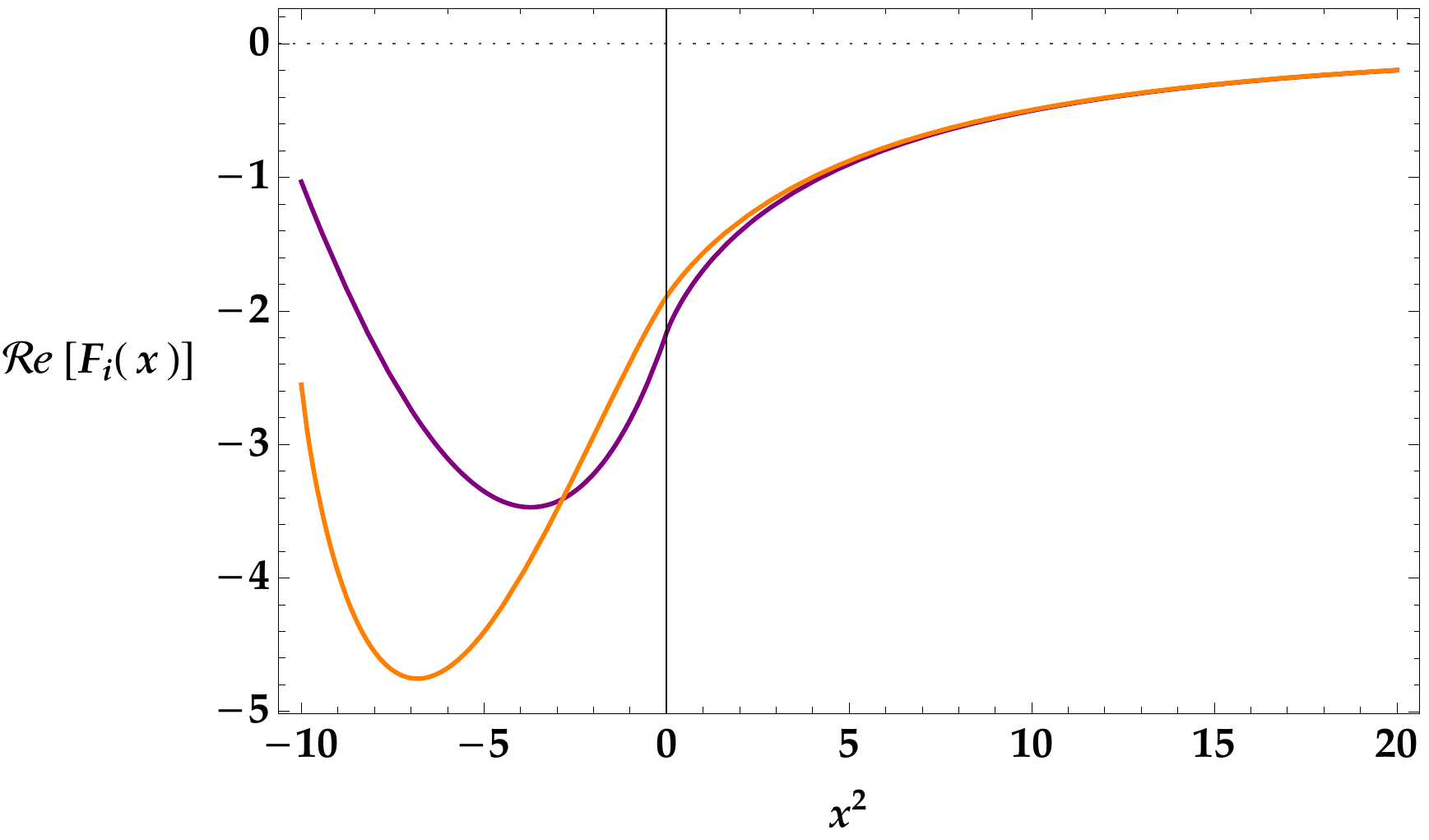}
  \caption{\label{fig:bf} Real parts of the bosonic thermal function $F_{b}(x)$ (purple curve) and the fermionic thermal function $F_{f}(x)$ (orange curve) are plotted over $x^2 = m^2/T^2$. When $x^2 \geq 0$, $F_{b}$ and $F_{f}$ are both real and the absolute values of $F_{b}$ and $F_{f}$ decrease monotonically with increasing $x^2$. Note that $F_{b}$ and $F_{f}$ generally become complex functions when $x^2 < 0$ ({\it i.e.}, as mass-squared turns negative).}
  \end{figure}

The one-loop thermal correction to the potential of the Higgs field at temperature $T$ has been found to be \cite{Laine2016,Quiros1999,KolbTurner, Carrington1992}
\be \label{DelV}
\Delta V_{T,\,i}(h,T) =  g_{i}\, \frac{T^4}{2\pi^2} \cdot F_{i} (m_{i},\,T) \,~,
\ee
where $g_{i}$ is the number of degrees of freedom associated with particle $i$, and the thermal functions $F_{i}$ for bosonic and fermionic loop of particle with mass $m$ at temperature $T$ are given by
\ba
F_{b}(m,\,T)  &&= \int_{0}^{\infty} dq \,q^2 \, \mathrm{ln}  \lb  1-\mathrm{exp} \lp - \sqrt{q^2 + \frac{m^2}{T^2} } ~\rp  \rb       ~, \label{boson}
\\
F_{f} (m,\,T) &&= -\int_{0}^{\infty} dq \,q^2 \, \mathrm{ln}  \lb  1 +\mathrm{exp} \lp - \sqrt{q^2 + \frac{m^2}{T^2} } ~\rp  \rb         \label{fermion}~,
\ea
respectively. Figure \ref{fig:bf} shows the thermal functions by plotting over $x^2=m^2/T^2$. The one-loop contribution of the Higgs field itself can be obtained by identifying $m^2$ in eq.~(\ref{boson}) with the field-dependent effective mass-squared
\be  \label{mh}
m_{H}(h)^2 = \frac{d^2V_{0}(h)}{dh^2} ~.
\ee
Although eq.~(\ref{DelV}) is typically known as thermal corrections to the effective potential, it actually describes the \emph{Helmholtz free energy density} $\mathcal{F}$ of an ensemble of quantum fields in the heat bath, where the relations to the energy density $\rho$ and entropy density $s$ of the system are \cite{KolbTurner}
\ba
\mathrm{energy~density}~~ \rho &&= \mathcal{F} + T s = \Delta V_{T}(h,\,T) + T s~,  \label{rho}
\\
\mathrm{entropy~density}~~ s &&= - \frac{\partial}{\partial T}\Delta V_{T}(h,\,T)~. \label{s}
\ea

In order to deal with the loop corrections in the model of Higgs inflation, special care with eqs.~(\ref{DelV}), (\ref{boson}), (\ref{fermion}) and (\ref{mh}) needs to be taken when considering the direct coupling between $h$ and the spacetime curvature. Essentially, the nonstandard term $\sim \xi R$ modifies the effective mass in eq.~(\ref{mh}) (see, for example, ref.~\cite{Hu1983}). The non-renormalizable coupling $\xi$  also results in additional corrections from graviton loop, making the theory much more complicated. On top of that, dynamics of the Higgs field and the cosmological observables for Higgs inflation are analyzed in the EF. In general, there are two possible choices to perform the quantum loop corrections for the potential:
\begin{itemize}
    \item  \emph{Prescription I} --- First transform to the EF and then compute quantum loop corrections with the EF potential in eq.~(\ref{EFU}).
    \item  \emph{Prescription II} --- First compute quantum loop corrections with the JF potential in eq.~(\ref{JVh}) and then transform to the EF.
\end{itemize}
The results of the two prescriptions for the radiative corrections at zero temperature up to two-loop level have been discussed in refs.~\cite{Allison2014,Bezrukov2009,Bezrukov2009-2,George2014}. Given the fact that relatively little is known about quantum gravity, there is no obvious preference for the prescription that one should take. Although the quantum scale invariance can support Prescription I \cite{Bezrukov2009-2}, one may also argue that Prescription II is more reliable since the physical distance scale is well-defined in the JF \cite{Barvinsky2008}. In this work, we shall adopt Prescription I for two reasons: (a) it can remove the uncertainty from graviton loops in the JF; and (b) the inflaton field $\chi$ ({\it i.e.}, the local ``clock'' of inflation) and its vacuum state, the Bunch-Davies vacuum, are all defined in the EF.  Therefore, it is more consistent to account for the finite-temperature loop corrections in the EF. In principle, we can assume that both prescriptions are equivalent at the fundamental level, since the physics should be the same under the conformal transformation. 

%==============
\subsection{Corrections from all degrees of freedom in the SM \label{32}}
%==============

Let us now write down the one-loop thermal correction from the Higgs field to the potential (\ref{EFU}). Defining the EF proper temperature (see eq.~(\ref{Appds})) 
\be \label{TE}
\widetilde{T} \propto \tilde{a}(\tau)^{-1} \propto f^{-\frac{1}{2}}\,a(t)^{-1} 
~,
\ee
and using eqs.~(\ref{DelV}), (\ref{boson}), (\ref{fermion}) and (\ref{mh}) in the EF, we obtain
\be   \label{EFDU}
\Delta U_{T,\,\mathrm{Higgs}}\lp\chi,\,\widetilde{T}\rp 
= 
\frac{\widetilde{T}^4}{2\pi^2} \cdot F_{b}\lp \widetilde{m}_{H}(\chi),\,\widetilde{T} \rp ~,
\ee
where, for $h \gg M_{P}/\xi$, the physical Higgs mass in the EF is determined by\footnote{It is easy to check that eq.~(\ref{EFm}) is equivalent to eq.~(3.2) in ref.~\cite{George2014}.}
\be \label{EFm}
\widetilde{m}_{H}(\chi)^2 
= 
\frac{d^2 U_{0}(\chi)}{d\chi^2} \approx \frac{d^2}{d\chi^2}\lv \frac{\lambda M^4_{P}}{4\,\xi^2}\,\lb 1-\mathrm{exp}\lp \frac{-2\,\chi}{\sqrt{6}M_{P}} \rp \rb^2 \rv  ~.
\ee
Besides loop corrections of the Higgs boson in the EF given in eq.~(\ref{EFm}), we should also take the Nambu-Goldstone (NG) bosons into account. The four degrees of freedom in the complex Higgs doublet $\mathcal{H}=(h_{1}+\mathrm{i}h_{2},\,h_{3}+\mathrm{i}h_{4})^{\mathrm{T}} /\sqrt{2}$ will mix under the non-canonical kinetic term $\sim \gamma_{ij} \partial h_{i} \partial h_{j} $ in the EF action \cite{George2014}, where 
\be \label{gamma}
\gamma_{ij} = \frac{1}{f} \lp \delta_{ij} + \frac{6\,\xi^2}{M_{P}^2 f}h_{i}h_{j} \rp  ~.
\ee
With the field metric in eq.~(\ref{gamma}), the associated mass-squared $\widetilde{m}_{G}^2$ in the EF is derived in ref.~\cite{George2014} for $\xi \gg 1$ via the covariant generalization of the second derivative of the potential
\be \label{MG}
\widetilde{m}_{G}(\chi)^2 
= \frac{\lambda 
\, h(\chi)^2}{f^2 \lb 1+6\,\xi^2 h(\chi)^2 / M_{P}^2\rb} 
\approx \frac{\lambda M_{P}^2}{6\,\xi^2} \, \mathrm{exp}\lp -\frac{4\,\chi}{\sqrt{6}M_{P}} \rp ~,
\ee
where we have used the approximation (\ref{hlarge}) in the second equality. Note that both eq.~(\ref{EFm}) and eq.~(\ref{MG}) are valid in the large-field (inflationary) regime. It follows that at the onset of inflation
\be
\frac{\widetilde{m}_{G}}{\left|\widetilde{m}_{H}\right|} \sim \mathrm{exp}\lp -\frac{\chi}{M_{P}} \rp \ll 1~,
\ee
so the NG modes are highly suppressed in the Coleman-Weinberg (CW) effective potential compared to the physical Higgs mode in the inflationary era. Nevertheless, we emphasize that we cannot ignore the one-loop thermal correction from the NG bosons. As we can see in figure \ref{fig:bf}, the absolute values of the thermal functions $F_{b}$ and $F_{f}$ at a fixed temperature increase as the particle mass decreases. In the high-temperature limit $\widetilde{T} \gg \widetilde{m}$, $F_{b}( \widetilde{m}_{G},\,\widetilde{T} )$ and $F_{b}( \widetilde{m}_{H},\,\widetilde{T} )$ will converge to comparable values. 

We also include other degrees of freedom in the SM (see table \ref{tab1}) to ensure the completeness for our numerical calculation in section \ref{4}. In the early Universe, all elementary particles: the Higgs boson, gauge bosons ($W^{\pm}$, $Z^{0}$), quarks ($u$, $d$, $c$, $s$, $t$, $b$), leptons ($e$, $\mu$, $\tau$, $\nu_{e}$, $\nu_{\mu}$, $\nu_{\tau}$), photon $\gamma$ and gluon $g$ could make contributions to the one-loop thermal effective potential. In the EF, the one-loop effective potential reads
\be \label{thermSM}
\Delta U_{T} \lp \chi,\,\widetilde{T}\rp = \frac{\widetilde{T}^4}{2\pi^2} \sum_{i} g_{i} F_{i} \lp \widetilde{m}_{i},\,\widetilde{T} \rp~,  
\ee
with
\ba
\sum_{i} g_{i} F_{i} \lp \widetilde{m}_{i},\,\widetilde{T} \rp && 
= 
\underbrace{F_{b} \lp \widetilde{m}_{H},\,\widetilde{T} \rp 
+ 3F_{b} \lp \widetilde{m}_{G},\,\widetilde{T} \rp}_{\displaystyle \mathrm{Higgs}} 
\\
&&
+ \underbrace{6F_{b}\lp\widetilde{m}_{W},\,\widetilde{T}\rp}_{\displaystyle \mathrm{W^{\pm}~boson}}
+\underbrace{3F_{b}\lp\widetilde{m}_{Z},\,\widetilde{T}\rp}_{\displaystyle \mathrm{Z^{0}~boson}} 
+ \underbrace{2 F_{b}\lp 0,\,\widetilde{T} \rp}_{\displaystyle \mathrm{photon}} 
+ \underbrace{16 F_{b}\lp 0,\,\widetilde{T} \rp}_{\displaystyle \mathrm{gluon}}  \nn
\\
&& 
+ \underbrace{12 \sum_{q}^{6} F_{f} \lp \widetilde{m}_{q},\,\widetilde{T} \rp }_{\displaystyle \mathrm{quark} }
+\underbrace{4 \sum_{\mathrm{cl}}^{3} F_{f} \lp \widetilde{m}_{\mathrm{cl}},\,\widetilde{T} \rp }_{\displaystyle \mathrm{charged~lepton} }
+\underbrace{2 \sum_{\nu}^{3} F_{f} \lp \widetilde{m}_{\nu},\,\widetilde{T} \rp }_{\displaystyle \mathrm{neutrino} }     ~.   \nn
\ea
Since we are computing the one-loop effective potential in the EF, all physical quantities must be correspondingly defined in the EF. Under transformation (\ref{CF0}), the mass $\widetilde{m}$ of the SM gauge bosons and fermions is scaled by
\be \label{CFm}
m_{i} ~~~~ \rightarrow ~~~~ \widetilde{m}_{i} = f(h)^{-1/2} \,m_{i}~
\ee
after transforming to the EF. See appendix \ref{A} for more detailed transformation rules.

\begin{table}[h!]
\centering
\begin{tabular}{c c c c c c}
\hline\hline
Species & Spins & With antiparticle & Colors & Flavors & d.o.f. \\
\hline 
Higgs $H^{0}$ & $1$ & $1$ & $1$ & $1$ & $1$ \\

gauge bosons $W^{\pm}$ & $3$ & $2$ & $1$ & $1$ & $6$ \\ 

$\qquad \qquad \quad~~ Z^{0}$ & $3$ & $1$ & $1$ & $1$ & $3$ \\ 
photon $\gamma$ & $2$ & $1$ & $1$ & $1$ & $2$ \\

gluon $g$ & $2$ & $1$ & $8$ & $1$ & $16$ \\

quarks $q$ & $2$ & $2$ & $3$ & $6$ & $72$ \\

charged leptons $l^{\pm}$ & $2$ & $2$ & $1$ & $3$ & $12$ \\

neutrinos $\nu$ & $1$ & $2$ & $1$ & $3$ & $6$ \\
\hline
\end{tabular}
\caption{\label{tab1} The degrees of freedom of all elementary particles in the SM.}
\end{table}

%==============
\subsection{The Coleman-Weinberg effective potential and the RG evolution} \label{33}
%==============
 
Since we are implementing the thermal corrections to the potential at the one-loop level, we must also include the CW effective potential in the EF. The CW correction for the Higgs inflation has been studied in refs.~\cite{Bezrukov2009, Bezrukov2009-2, Allison2014}, and is approximated by 
\ba \label{UCW}
\Delta U_{\mathrm{CW}}
&& \approx \frac{1}{16\pi^2} \lv 
 \frac{\widetilde{m}_{H}^4}{4} \, \lb \mathrm{ln} \lp \frac{\widetilde{m}_{H}^2}{\widetilde{\mu}^2}\rp -\frac{3}{2} \rb  + \frac{3\widetilde{m}_{G}^4}{4} \, \lb \mathrm{ln} \lp \frac{\widetilde{m}_{G}^2}{\widetilde{\mu}^2}\rp -\frac{3}{2} \rb  \right. 
\\
&& + \frac{3\widetilde{m}_{W}^4}{2} \,\lb \mathrm{ln} \lp \frac{\widetilde{m}_{W}^2}{\widetilde{\mu}^2}\rp-\frac{5}{6}\rb +\frac{3\widetilde{m}_{Z}^4}{4} \, \lb \mathrm{ln} \lp \frac{\widetilde{m}_{Z}^2}{\widetilde{\mu}^2}\rp - \frac{5}{6} \rb  
\left. -3\widetilde{m}_{t}^4 \, \lb \mathrm{ln} \lp \frac{\widetilde{m}_{t}^2}{\widetilde{\mu}^2}\rp -\frac{3}{2} \rb ~\rv ~,\nn
\ea    
with $\widetilde{\mu}$ being the EF renormalization scale. As described in section \ref{31}, we can use either Prescription I or Prescription II to define the renormolization scale. We again adopt Prescription I so that the quantum corrections are calculated in the EF with mass parameters $\widetilde{m}=f^{-1/2}\, m$ and  $\widetilde{\mu} \sim f^{-1/2} \, h(\chi)$ \cite{Bezrukov2009, Bezrukov2009-2, Allison2014, Hamada2017}. The complete finite-temperature potential then reads
\be \label{Utotal}
\mathcal{U} \lp \chi,\widetilde{T}\rp  
=  U_{0}(\chi) + \Delta U_{\mathrm{CW}} \lp \chi \rp + \Delta U_{T} \lp\chi,\widetilde{T}\rp = U_{0}(\chi) + \Delta U_{\mathrm{eff}}\lp\chi,\widetilde{T}\rp~.
\ee
However, eq.~(\ref{CF5}) tells us that the mass parameters of the SM particles $\widetilde{m}_{W}$, $\widetilde{m}_{Z}$ and $\widetilde{m}_{t}$ are approximately proportional to $f^{-1/2} \,y \,h(\chi) \sim y\,M_{P}/\sqrt{\xi} $, where $y$ represents either gauge or Yukawa couplings. As a result, the CW effective potential in eq.~(\ref{UCW}) is generally small compared to the plateau of the tree-level potential during inflation
\be
\Delta U_{\mathrm{CW}} \sim \frac{1}{16\pi^2}\widetilde{m}^4 \sim \frac{1}{16\pi^2} \frac{y^4 M_{P}^4}{\xi^2} ~\lesssim~ \mathcal{O} \lp10^{-2}\rp \cdot \frac{\lambda M_{P}^4}{4\,\xi^2} \sim \mathcal{O} \lp 10^{-2}\rp \cdot U_{0}(\chi)    ~.   \nn
\ee
It means that one can safely ignore the CW term in eq.~(\ref{Utotal}) in analytic calculations.

In this work, we do not consider the renormalization group (RG) running of the SM coupling constants and the field-dependent suppression factor of RG equations discussed in refs.~\cite{Elizalde1993,Elizalde1994,Elizalde1994-2,Allison2014,Simone2009}. In fact, our results presented in section \ref{4} would not be sensitive to any expected changes of the coupling constants. As we can see in section \ref{43}, most SM particles are ultra-relativistic for the temperatures we are interested in ({\it i.e.}, $\widetilde{T} \gg \widetilde{m}$); hence, the exact values of the mass parameters (which can vary with RG running) are not important to the thermal effective potential. Regarding the running of the Higgs self-coupling $\lambda$, we would like to emphasize that the whole scenario of the Higgs inflation does not explicitly depend on the value of the quartic coupling $\lambda$.  Instead, all of the important calculations and derived quantities depend on the combination $\lambda/\xi^2$ ({\it i.e.}, $\lambda$ is always associated with $\xi^2$ in the denominator). For example, the tree-level potential $U_0 (\chi)$ of the Einstein frame defined in eq.~(\ref{Ularge}), the Higgs mass-squared $\widetilde{m}^2_H = U_0''(\chi)$ in eq.~(\ref{EFm}), the Nambu-Goldstone boson mass-squared $\widetilde{m}^2_G$ in eq.~(\ref{MG}), and the one-loop thermal correction $\Delta U_T (\chi,\,\widetilde{T} ) $ defined in eq.~(\ref{EFDU}) all contain direct dependence on $\lambda/\xi^2$.  Therefore, our results in section \ref{4} are only sensitive to $\lambda/\xi^2$ rather than the individual values of $\lambda$ and $\xi$ when most SM particles are relativistic. On the other hand, certain cosmological parameters such as the potential slow-roll parameters in eqs.~(\ref{epsilonHiggs}) and (\ref{etaHiggs}) are independent of $\lambda/\xi^2$.\footnote{Note that $h$ is always scaled by $M_P/\sqrt{\xi}$ for inflation, so the potential slow-roll parameters are also independent of $\xi$.} Given the WMAP normalization in eq.~(\ref{WMAPNorm}), we can directly constrain $\lambda/\xi^2$ to a fixed value during inflation. As a result, any variation in $\lambda$ can always be compensated by a change in $\xi$. In other words, the uncertainties of the running in quartic coupling $\lambda$ are not important to our final results.\footnote{A potential problem could arise if $\lambda \rightarrow 0 $. When $\lambda \sim 10^{-9}$, the corresponding $\xi$ is of order unity due to CMB constraint (\ref{WMAPNorm}). In this case, we would need the Higgs field $h \sim M_P/\sqrt{\xi}$ to reach the trans-Planckian scale for inflation and non-perturbative effects may spoil the flatness of potential. The issue basically arises as $\xi < \mathcal{O}(10^1)$, which corresponds to $\lambda < \lambda_{\mathrm{min}} \sim \mathcal{O}(10^{-7}) $. However, as pointed out by refs.~\cite{Allison2014, Simone2009}, the existence of the field at trans-Planckian scale is a general problem appearing in many minimal models of inflation rather than just for Higgs inflation.} This argument has also been noted in the literature, such as refs.~\cite{Bezrukov2015, Rubio2019, George2014, Allison2014, Simone2009}.

For the issue of the Higgs criticality at high scales, earlier works such as refs.~\cite{Bezrukov2015,Rubio2019} point out that, in the scenario of the Higgs inflation, the self-coupling $\lambda$ can be brought to positive value by taking the additional renormalization effects at the scale $\chi \sim M_{P}/\xi$ into account. This provides the theoretical basis that $\lambda$ can remain positive at the inflationary scale (which is well above $M_{P}/\xi$) and successful Higgs inflation can still take place even the instability due to the Higgs self-coupling happens at certain scale $\widetilde{\mu}_0$ (which is typically below $M_{P}/\xi$). Furthermore, the asymptotic shape of the effective potential involving both the Higgs instability and the renormalization in the inflationary regime would coincide with the simple scenario we have studied \cite{Bezrukov2015}. All in all, the Higgs inflation can still be regarded as a viable model even when the running of $\lambda$ is taken into account and its RG running is largely irrelevant. The arguments above together support our implicit assumption that including the RG running of the SM coupling constants is not crucial to our main results.

%==============
\subsection{Quantum instability} \label{34}
%==============

 According to eq.~(\ref{Ularge}), the tree-level potential of $\chi$ is convex during the inflation. It follows that the Higgs mass-squared $\widetilde{m}_{H}^2 = U_{0}''(\chi)$ is negative in the EF. Both the finite-temperature effective potential defined by eq.~(\ref{boson}) and the CW effective potential in eq.~(\ref{UCW}) would become complex, signaling a quantum instability of the $\chi$ field \cite{Weinberg2015}. It has been argued in ref.~\cite{Weinberg1987} that the imaginary part of the one-loop perturbative effective potential $\Delta U_{\mathrm{CW}}$ is physically related to the decay rate per unit volume of the unstable state.\footnote{Ref.~\cite{Weinberg1987} studied a complex CW effective potential. However, the physical interpretation of the imaginary part of the effective potential, an off-shell quantity, remains somewhat controversial. Here we assume that the imaginary part of the thermal effective potential $\Delta U_{T}$ is unphysical and only the real part is relevant to our calculations.} Fortunately, we will find that the decay process is generally insignificant during the inflation. Given a finite temperature $\widetilde{T}$ in the inflationary era, the average energy density of $\chi$ is enhanced by the background temperature, and the expansion rate $\widetilde{H}_{\mathrm{inf}}$ of the Universe becomes larger. By the end of inflation, $\widetilde{T}\rightarrow 0$ and the expansion rate $\widetilde{H}_{\mathrm{end}}$ is suppressed. Consequently, we have
 $\widetilde{H}_{\mathrm{end}} < \widetilde{H}_{\mathrm{inf}}$ and the decay rate of $\chi$ during the inflation 
 \ba \label{Decay}
 \Gamma_{\mathrm{inf}} 
 && \sim \mathrm{Im}\lp\Delta U_{\mathrm{CW}} \rp \cdot \widetilde{H}_{\mathrm{inf}}^{-3} ~<~ \mathrm{Im}\lp\Delta U_{\mathrm{CW}} \rp \cdot \widetilde{H}_{\mathrm{end}}^{-3} \nn
 \\
 &&\sim \mathrm{Im}\lp\Delta U_{\mathrm{CW}} \rp \cdot \lv \frac{M_{P} \sqrt{\lambda}}{2\sqrt{3}\,\xi}\lb 1-\mathrm{exp}\lp \frac{-2\,\chi_{\mathrm{end}}}{\sqrt{6}M_{P}} \rp \rb \rv^{-3} ~. 
 \ea
Using $\chi_{\mathrm{end}} = 0.94~ M_{P}$, $\xi/\sqrt{\lambda} = 4.7 \times 10^4 $ (see section \ref{2}), we numerically estimate $\mathrm{Im}\lp \Delta U_{\mathrm{CW}}\rp$ and find
\be  \label{Decay2}
\Gamma_{\mathrm{inf}} \ll \widetilde{H}_{\mathrm{end}} < \widetilde{H}_{\mathrm{inf}} ~.
\ee
Eq.~(\ref{Decay2}) suggests that the decay rate of $\chi$ is negligible comparing to the expansion rate of the Universe during the inflation. Along with the fact that the free energy should be real, we will only consider the real part of the effective potential for the numerical calculations in section \ref{4}.

%=================================================================%
\section{Thermal effects on CMB: the primordial curvature power spectrum} \label{4}
%=================================================================%

Through the large Hubble friction in the early Universe, the average kinetic energy of $\chi$ is maintained to be much smaller than the potential energy, leading to a quasi-de Sitter phase of the cosmic expansion. The quantum fluctuations at large scales are then gradually frozen ({\it i.e.}, not in causal contact) due to the shrinking comoving horizon, leaving a scale-invariant primordial curvature power spectrum $P_{\mathcal{R}}(k)$. Together with the composition of the Universe after reheating and the late-time cosmological processes, we can reconstruct the angular power spectra $\mathcal{D}^{TT}_{\ell}$ of the CMB and compare it with that determined by the correlation function of the CMB temperature anisotropies. Investigating the primordial power spectrum thus facilitates the constraints on the physics of inflation \cite{Sasaki1986,Mukhanov1992,Martin2014,Planck2018const,Leach2003,Hannestad2001,Zeng2019}. In this section, we examine how the formalism of the finite-temperature loop corrections in the EF discussed in section \ref{3} can affect the dynamics of the inflaton $\chi$ and the primordial power spectrum of the quantum perturbation during the inflation. The issue of whether thermal equilibrium between all SM particles can be established will be discussed at the end of this section.

%==============
\subsection{The field equations and the power spectrum}\label{41}
%==============

Suppose that all the quantum fields in the Universe are in thermal equilibrium with a heat bath of temperature $\widetilde{T}$ before the inflation happens, the energy density associated with $\chi$ in the EF is \cite{KolbTurner}
\be
\rho = \mathcal{U}\lp\chi,\,\widetilde{T}\rp +\widetilde{T}\, \tilde{s}   ~, \label{thermrhoE}
\ee
where
\be
\tilde{s} = - \frac{\partial}{\partial \widetilde{T}}\,\Delta U_{T}\lp\chi,\,\widetilde{T}\rp   \label{sE}
\ee
and the temperature-dependent potential $\mathcal{U}$ is given in eq.~(\ref{Utotal}). Additionally, the second law of thermodynamics gives
\be \label{thermo2}
\tilde{s} = \frac{ \rho + \mathcal{P}  }{\widetilde{T}}~.
\ee
Using eqs.~(\ref{thermrhoE}) and (\ref{thermo2}), we immediately obtain the pressure
\be\label{thermpE}
\mathcal{P}  =   -\mathcal{U} \lp\chi,\,\widetilde{T}\rp~. 
\ee
Typically, the kinetic energy associated with the microscopic fluctuations $\delta \chi$ around the homogeneous field expectation value $\left \langle \chi \right \rangle \equiv \chi_{c}$ is included in the finite-temperature effective potential $\Delta U_{T}$. But we are now focusing on the macroscopic dynamics of $\chi_{c}$ under the influence of the inflationary potential and the Hubble friction, the expectation value $\chi_{c}$ is in general time-dependent. Therefore, we may consider the \emph{drift velocity} of $\chi_{c}$ by rewriting eqs.~(\ref{thermrhoE}) and (\ref{thermpE}) as\footnote{A decomposition of the field by $\chi = \left \langle \chi \right \rangle + \delta \chi$ yields $\dot{\chi}^2  = \dot{\left \langle \chi \right \rangle}^2 + 2 \dot{\left \langle \chi \right \rangle}\, \delta\dot{\chi}  + \delta\dot{\chi}^2 \approx \dot{\chi_{c}}^2 + \delta\dot{\chi}^2  $ on average. The former part corresponds to the kinetic energy with drift velocity while the latter part is the kinetic energy associated with random fluctuations.}
\ba
\rho_{*} = &&   \frac{1}{2}\lp\frac{d\chi_{c}}{d\tau}\rp^2  + \mathcal{U}\lp\chi,\,\widetilde{T}\rp + \widetilde{T}\,\tilde{s} ~;   \label{thermrhoE2}  
\\
\mathcal{P}_{*} =&&\frac{1}{2}\lp\frac{d\chi_{c}}{d\tau}\rp^2  - \mathcal{U}\lp\chi,\,\widetilde{T}\rp~.  \label{thermpE2} 
\ea
With eqs.~(\ref{thermrhoE2}) and (\ref{thermpE2}) and the equation of energy conservation,
we obtain
\be
\frac{d\chi_{c}}{d\tau} \cdot  \frac{d^2\chi_{c}}{d\tau^2}  +  \frac{d\chi_{c}}{d\tau}~ \cdot  \frac{d\rho_{*} (\chi,\,\widetilde{T})}{d\chi_{c}} = - 3 \widetilde{H} \lb \lp\frac{d\chi_{c}}{d\tau}\rp^2 + \widetilde{T}\, \tilde{s}\, \rb~.\nn
\ee
Treating $\chi_{c}$ as a thermodynamic variable and expanding the total derivative, the equation of motion for $\chi_{c}$ is
\be\label{EOMchi2}
\frac{d^2\chi_{c}}{d\tau^2} + 3\widetilde{H} \lp \frac{d\chi_{c}}{d\tau}+  \frac{\widetilde{T}\,\tilde{s}}{d\chi_{c}/d\tau}  \rp  - \widetilde{T} \cdot \frac{\partial^2 \Delta U_{T}}{\partial\, \widetilde{T}^2} \cdot \frac{d \widetilde{T}}{d\chi_{c}} = - \frac{\partial}{\partial\chi_{c}}  \lb \,\mathcal{U}\lp\chi,\,\widetilde{T}\rp + \widetilde{T}\,\tilde{s} \,\rb~.     
\ee
In the zero-temperature limit $\widetilde{T} \rightarrow 0$ at tree level, eq.~(\ref{EOMchi2}) recovers the classical inflaton field equation
\be \label{EOMinf}
\frac{d^2\chi_{c}}{d\tau^2} + 3\widetilde{H} \,\frac{d\chi_{c}}{d\tau} = - \frac{d}{d\chi_{c}}U_{0}(\chi_{c})~.
\ee
In addition, the expansion rate of the Universe is determined by the Friedmann equation
\be \label{FE}
\widetilde{H}^2 = \frac{1}{3M_{P}^2} \,\rho_{*} ~.
\ee
To avert clutter, we will represent the derivative with respect to $\tau$ by a ``dot'' henceforth. Eqs.~(\ref{EOMchi2}) and (\ref{FE}) together describe a complete dynamical system in the EF. However, when adding up all degrees of freedom in the SM, it is formidable to perform exact numerical calculations. One way to overcome this difficulty is expanding the thermal functions in eqs.~(\ref{boson}) and (\ref{fermion}) into an infinite series of the modified Bessel function of the second kind
\ba
F_{b}\lp\, \widetilde{m},\,\widetilde{T}\, \rp && = - \frac{\widetilde{m}^2}{\widetilde{T}^2}\cdot \sum_{n=1}^{\infty}\frac{1}{n^2}K_{2}\lp n\cdot \frac{\widetilde{m}}{\widetilde{T}} \rp~, \label{K2boson}
\\
F_{f}\lp\, \widetilde{m},\,\widetilde{T} \,\rp && = \frac{\widetilde{m}^2}{\widetilde{T}^2}\cdot \sum_{n=1}^{\infty}\frac{(-1)^n}{n^2}K_{2}\lp n\cdot \frac{\widetilde{m}}{\widetilde{T}} \rp~. \label{K2fermion}
\ea
We find that sufficient precision and faster calculations can be achieved with around $10$ terms in eqs.~(\ref{K2boson}) and (\ref{K2fermion}). 

Provided the initial condition $\dot{\chi_{c}}^2 \ll \mathcal{U}$ during the inflation, the slow-roll approximation is applicable and the primordial curvature power spectrum at the hoziaon crossing at the scale $k^{-1} = \widetilde{H}^{-1}/\tilde{a}$ can be written as
\be \label{Pk}
P_{\mathcal{R}} (k) = \left.\frac{1}{8\pi^2 M_{P}^2}\frac{\widetilde{H}^2}{\varepsilon}\right|_{k=\tilde{a}\widetilde{H}}~~,~~~~\varepsilon = - \frac{\dot{\widetilde{H}}}{\widetilde{H}^2}~,
\ee
where $\varepsilon$ is the Hubble slow-roll parameter. Eq.~(\ref{Pk}) is defined by the variance of the quantum fluctuations with respect to the Bunch-Davies vacuum. The finite-temperature effect of the pre-inflationary radiation field on the primordial power spectrum is also studied in refs.~\cite{Bhattacharya2006,Powell2007,Wang2008,Das2015} by considering either non-trivial mode functions of the primordial perturbation or a thermal distribution of the inflaton, which will modify eq.~(\ref{Pk}) with additional temperature-dependent factors and may lead to a different choice of the vacuum state. Here we take into account the thermal effect from the first principle by computing the quantum loops in the Bunch-Davies vacuum.

%==============
\subsection{High-temperature approximations}\label{42}
%==============

Solving eqs.~(\ref{EOMchi2}) and (\ref{FE}) is a thorny task. Fortunately, we do not need to bother with exact solutions of the set of differential equations. To see how the thermal quantum loops affect the power spectrum, we can first use eqs.~(\ref{thermrhoE2}), (\ref{thermpE2}) and (\ref{FE}) to obtain
\be
\dot{\widetilde{H}}  = \frac{\ddot{\tilde{a}}}{\tilde{a}} - \widetilde{H}^2 = -\frac{1}{2M_{P}^2}\lp \rho_{*}+\mathcal{P}_{*} \rp 
= -\frac{1}{2M_{P}^2}\lp \dot{\chi_{c}}^2 + \widetilde{T}\,\tilde{s} \rp ~. 
\ee
Eq.~(\ref{Pk}) then gives
\be 
\varepsilon  = \frac{3}{2}\frac{\dot{\chi_{c}}^2+\widetilde{T}\, \tilde{s} }{\dot{\chi_{c}}^2/2 + \mathcal{U} + \widetilde{T} \,\tilde{s} } 
\ee
and
\be \label{Pk2}
P_{\mathcal{R}} =  \frac{1}{36 \pi^2 M_{P}^4} \frac{\lp \dot{\chi_{c}}^2/2 + \mathcal{U} + \widetilde{T}\,\tilde{s}\rp^2 }{  \dot{\chi_{c}}^2 + \widetilde{T}\,\tilde{s} }  ~.
\ee
We have discussed in section \ref{33} that the CW correction $\Delta U_{\mathrm{CW}}$ is fairly small compared to the tree-level potential $U_{0}$ in the inflationary era, and will thus neglect it in the following approximations. Also, we can Taylor-expand the thermal functions (\ref{boson}) and (\ref{fermion}) about $ x = \widetilde{m}/\widetilde{T} = 0$ in the high-temperature limit $\widetilde{T}\gg \widetilde{m}$ as
\ba
F_{b} (x) && = -\frac{\pi^4}{45} + \frac{\pi^2}{12}x^2 -  \frac{\pi}{6}x^3 - \frac{x^4}{32}\, \mathrm{ln}\lp\frac{x^2}{a_{b}} \rp +~ \cdot\cdot\cdot~; \label{boson2}
\\
F_{f} (x) && = -\frac{7\pi^4}{360} + \frac{\pi^2}{24}x^2 + \frac{x^4}{32}\, \mathrm{ln}\lp\frac{x^2}{a_{f}} \rp +~ \cdot\cdot\cdot~, \label{fermion2}
\ea
with $a_{b}=3/2 - 2\gamma_{e} + 2\,\mathrm{ln}(4\pi)$, $a_{f}=3/2 - 2\gamma_{e} + 2\,\mathrm{ln}(\pi)$ and the Euler-Mascheroni constant $\gamma_{e} \simeq 0.577$. It is then useful to approximate eq.~(\ref{thermSM}) by
\be \label{thermSMhighT}
\Delta U_{T} \approx  -\lp  \frac{\pi^2}{90} \sum_{b} g_{b} + \frac{7\pi^2}{720} \sum_{f}g_{f} \rp\widetilde{T}^4~~~~\mbox{for}~ \widetilde{T} \gg \widetilde{m}~.
\ee
The dependence on mass parameters $\widetilde{m}$ disappears in eq.~(\ref{thermSMhighT}) and there is a $\widetilde{T}^4$ dependence in the formula. This is because all particles become ultra-relativistic and behave like massless radiation in the high-temperature limit. Together with the slow-roll approximation, we can further rewrite eq.~(\ref{Pk2}) as
\be
P_{\mathcal{R}}  \approx  
\frac{5}{4\pi^4 M_{P}^4} \lb U_{0}  - \widetilde{T}^2 \frac{\partial}{\partial \widetilde{T}} \lp\frac{\Delta U_{T}}{\widetilde{T}}\rp \rb^2 \lp  2 \sum_{b}g_{b} + \frac{7}{4}\sum_{f}g_{f} \rp^{-1}  \widetilde{T}^{-4}~,  \label{Pk3}
\ee
where
\be\label{Ts4}
- \widetilde{T}^2 \frac{\partial}{\partial \widetilde{T}} \lp\frac{\Delta U_{T}}{\widetilde{T}}\rp \approx  \lp \frac{\pi^2}{30} \sum_{b} g_{b} + \frac{7\pi^2}{240} \sum_{f}g_{f} \rp \widetilde{T}^4 ~>~ 0 ~.
\ee

%==============
\subsection{Results}\label{43}
%==============

\begin{figure}[ht] 
  \centering
  \includegraphics[width=1\textwidth]{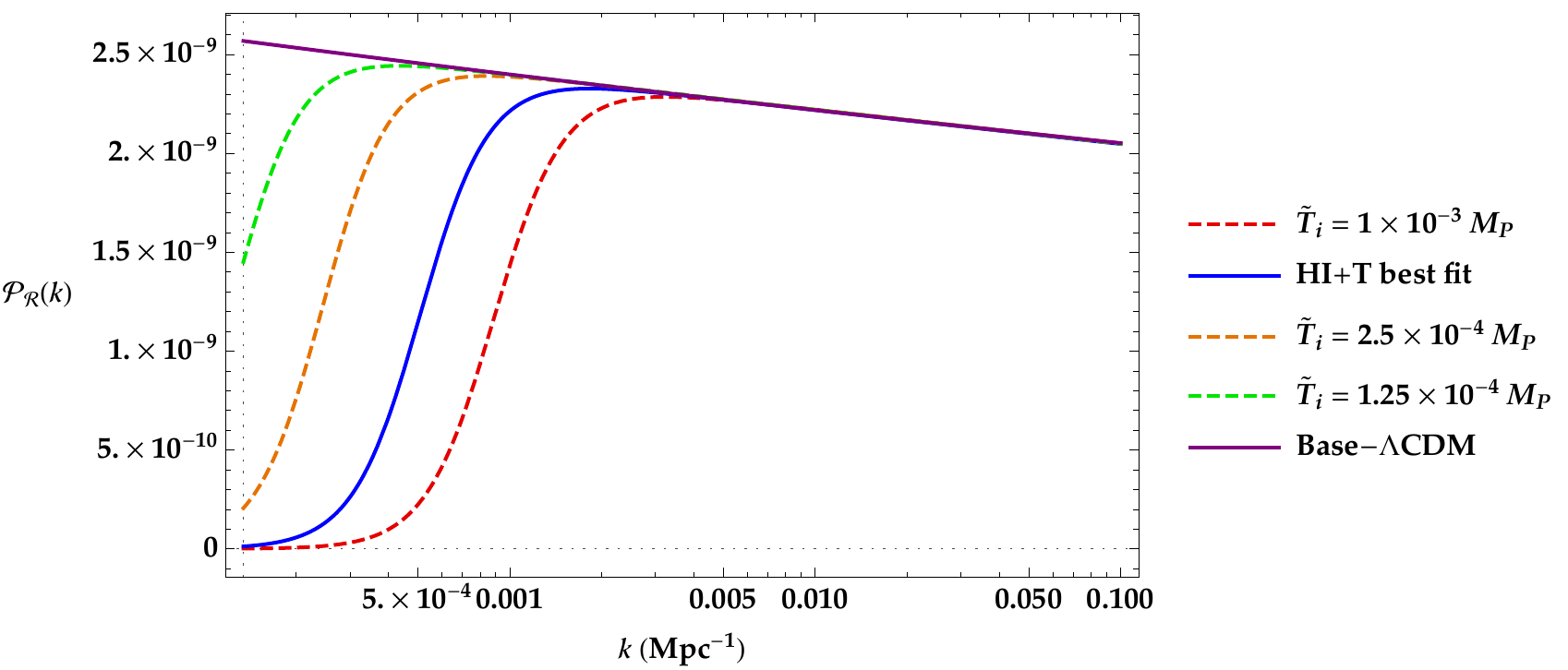}
  \caption{\label{fig:Pk} The primordial curvature power spectra $P_{\mathcal{R}}(k)$ for various initial temperatures $\widetilde{T}_{i}$. The standard power-law spectrum of the base-$\Lambda$CDM model, $P_{\mathcal{R}}(k )\propto k^{n_{s}-1}$, is given by the purple solid curve. The blue solid curve depicts the best-fit prediction of the Higgs inflation with an initial temperature $\widetilde{T}_{i}=5.15\times 10^{-4} ~M_{P}$ (or $3.62\times 10^{-3} ~M_{P}$).}
  \end{figure}

It turns out that the temperature dependence of $P_{\mathcal{R}}(k)$ is quite simple when $\widetilde{T}\gg \widetilde{m}$. For the case of extremely high temperature $\widetilde{T}^4 \gg U_{0}$, the second term in the square bracket of eq.~(\ref{Pk3}) may dominate over the potential plateau $U_{0}$, leading to $P_{\mathcal{R}}(k) \propto \widetilde{T}(k)^4$. In this regime, the amplitude of the power spectrum is enhanced by the thermal effects. On the other hand, eq.~(\ref{Pk3}) also indicates that the amplitude of the power spectrum can be suppressed by $\widetilde{T}(k)^{-4}$ if $\chi$ has an ultra-high temperature satisfying $\widetilde{T} \gg \widetilde{m}$ and $\widetilde{T}^4 \ll U_{0}$:
\be \label{Pk4}
P_{\mathcal{R}}(k) \approx  \left.\frac{5}{4\pi^4 M_{P}^4} U_{0}(\chi)^2 \lp  2 \sum_{b}g_{b} + \frac{7}{4}\sum_{f}g_{f} \rp^{-1}  \widetilde{T}(k)^{-4} \,\right|_{k = \tilde{a}\widetilde{H}} ~.
\ee
In the intermediate regime where $\widetilde{T}^4\sim U_{0}$, the qualitative behavior of the primordial power spectrum smoothly transits from $\widetilde{T}^4$ to $\widetilde{T}^{-4}$ as the temperature decreases. An order-of-magnitude estimation shows that the intermediate regime happens when $\widetilde{T} \simeq 10^{-3}~M_{P}$; that is, $P_{\mathcal{R}}(k)$ is enhanced with temperature if $\widetilde{T}>10^{-3}~M_{P}$ and suppressed if $\widetilde{T} \lesssim 10^{-3}~M_{P}$. We show in figure \ref{fig:Pk} the primordial curvature power spectra with various $\widetilde{T}_{i}$ in the ultra-high temperature regime ($\widetilde{T}_{i} \lesssim 10^{-3}~M_{P} $) when the CMB scale $k_{\mathrm{CMB}}^{-1}$ leaves the horizon.\footnote{In order to focus on the finite-temperature effects, we treat the number of the SM degrees of freedom as a constant for the results here.} As we increase the initial temperature, the magnitude of the spectrum is indeed considerably lowered. With an initial temperature $\widetilde{T}_{i} = 10^{-3} ~M_{P}$, $P_{\mathcal{R}}(k_{\mathrm{CMB}})$ is suppressed by more than three orders of magnitude comparing to the standard power-law predictions. Note that in the extremely-high temperature regime ($\widetilde{T}_{i} > 10^{-3} ~M_{P}$), $P_{\mathcal{R}}(k_{\mathrm{CMB}})$ is enhanced relative to the maximal suppression at $\widetilde{T} \simeq 10^{-3}~M_{P}$, so each power spectrum curve for $\widetilde{T}_{i} > 10^{-3} ~M_{P}$ can correspondingly find an almost identical one for $\widetilde{T}_{i} \lesssim 10^{-3} ~M_{P}$. Therefore, we only depict the curves for $\widetilde{T}_{i} < 10^{-3} ~M_{P}$ in figure \ref{fig:Pk}. Also, the numerical results in figure \ref{fig:Pk} do not exactly follow the high-temperature approximation in eq.~(\ref{Pk4}), since the mass parameters of the gauge bosons and quarks are proportional to $ f^{-1/2}\, h \sim M_{P}/\sqrt{\xi}  $ in the EF during the inflation. The masses of some heavy particles such as $\widetilde{m}_{W},\,\widetilde{m}_{Z},\,\widetilde{m}_{t},\,\widetilde{m}_{b}$ can be comparable to the initial temperature $\widetilde{T}_{i}$. 

In ref.~\cite{Das2015}, the authors have shown that the thermal initial state can enhance the power spectrum at large angular scales; however, the enhancement is dwarfed by the modification of the mode function from the pre-inflationary thermal bath. We compare our results in figure \ref{fig:Pk} with those in figure $1$ of ref.~\cite{Das2015}. It is apparent that for our model with the best-fit temperature (the blue solid curve in figure \ref{fig:Pk}), the power spectrum is further lowered by the one-loop thermal corrections of the inflation potential at the CMB scales, so considering the thermal vacuum state instead of the Bunch-Davies vacuum state would not substantially change our results. As such, our result for the power spectrum will give rise to a more pronounced power suppression in the CMB large-scale anisotropy (see the blue solid curve in figure \ref{fig:TT} below), as compared to that in figure $3$ of ref.~\cite{Das2015}. 
 
  \begin{figure}[ht] 
  \centering
  \includegraphics[width=1\textwidth]{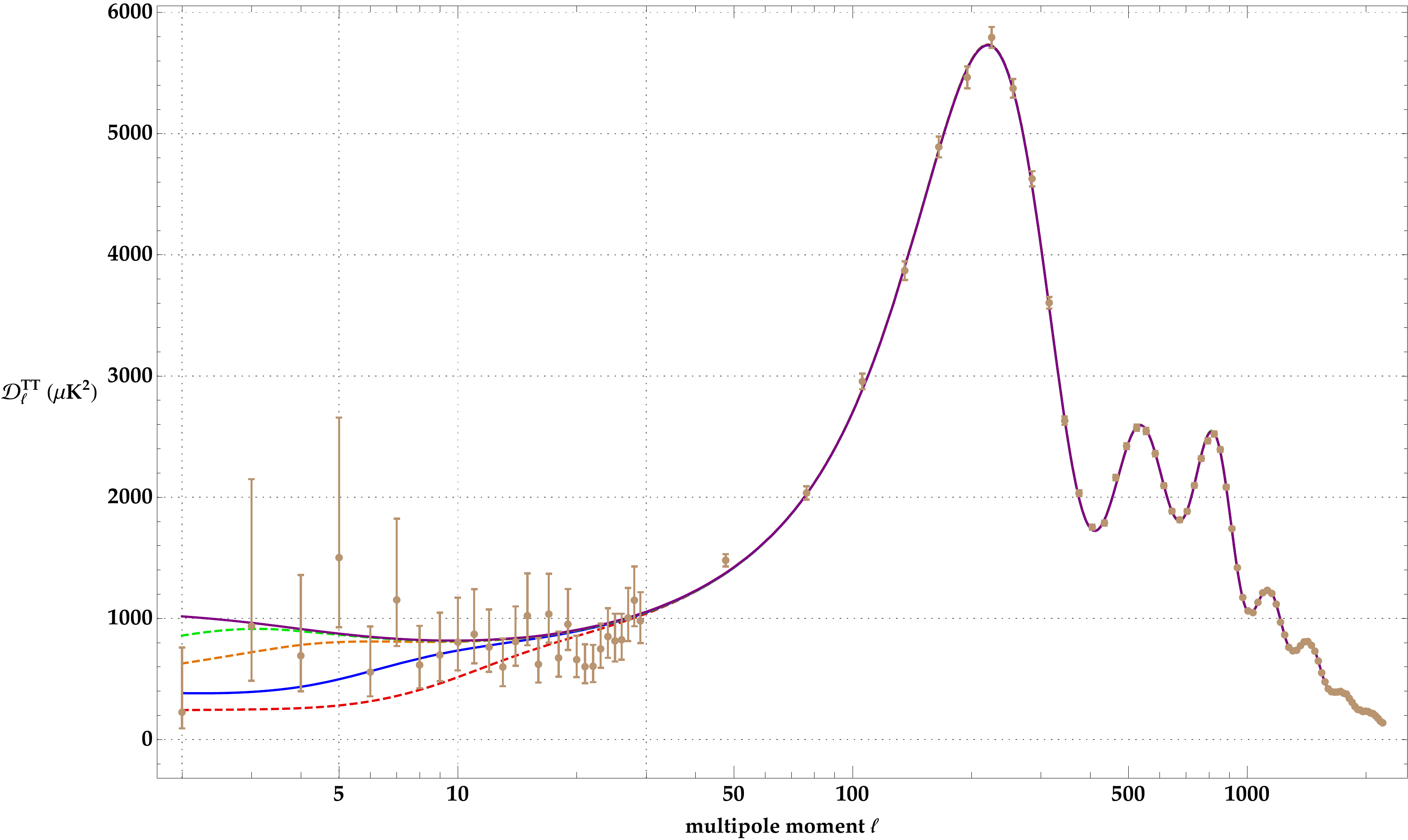}
  \caption{\label{fig:TT} The $TT$ angular power spectrum for the Higgs inflation with various initial temperatures $\widetilde{T}_{i}$ at the CMB horizon scale. The best-fit base-$\Lambda$CDM model is given by the purple solid curve.  Predictions of our model for various $\widetilde{T}_{i}$ are given by the red dashed curve ($1\times 10^{-3} ~M_{P}$), orange dashed curve ($2.5\times 10^{-4} ~M_{P}$), and green dashed curve ($1.25\times 10^{-4} ~M_{P}$). The blue solid curve depicts the best-fit prediction of the Higgs inflation with an initial temperature $\widetilde{T}_{i}=5.15\times 10^{-4} ~M_{P}$ (or $3.62\times 10^{-3} ~M_{P}$).}
  \end{figure}

Using the public Boltzmann code \texttt{CAMB} \cite{CAMB}, we further compute the $TT$ angular power spectra of the CMB based on the primordial power spectrum $P_{\mathcal{R}}(k)$. The results are demonstrated in figure \ref{fig:TT}. In order to fit the model to the Planck data \cite{Planck}, we calculate the values of reduced $\chi^2$ for $\ell = 2 \sim 29$ with the following formula
\be\label{chisquare}
\chi_{\mathrm{theory}}^2 = \frac{1}{28-1}\sum^{29}_{\ell=2} \frac{\lp\mathcal{D}^{TT}_{\ell,\,\mathrm{theory}}-\mathcal{D}^{TT}_{\ell,\,\mathrm{data}}\rp^2}{\sigma_{\ell,\,\mathrm{data}}^2}~.
\ee
The $\chi^2$ values of the models with different initial temperatures are listed in table~\ref{tab2}. From the minimum of the $\chi^2$ value, the best-fit initial temperatures in the ultra and extremely-high temperature regimes are found to be $\widetilde{T}_{i} \simeq 5.15 \times 10^{-4}~M_{P}$ and $\widetilde{T}_{i} \simeq 3.62 \times 10^{-3}~M_{P}$, respectively. In general, the thermal effects from quantum loop corrections to the Higgs inflation provide better fits to the current CMB angular power spectrum. This is due to the fact that the central values of many measurements $\mathcal{D}^{TT}_{\ell,\,\mathrm{data}}$ at $\ell < 30 $ are smaller than the best-fit base-$\Lambda$CDM model. Among the multipoles in the angular power spectrum, the low amplitude of the quadrupole moment $(\ell=2)$ is particularly significant. As shown in figure~\ref{fig:TT}, the Higgs inflation with thermal effects can be served as a possible solution to explain the anomaly with a single parameter, the initial $\widetilde{T}_{i}$ at the CMB horizon scale.

  \begin{table}[ht]
  \centering
  \begin{tabular}{c c c}
  \hline\hline
  \textbf{Model} & \textbf{Temperature at $k_{\mathrm{CMB}}$} $(M_{P})$  & \textbf{$\chi_{\mathrm{theory}}^2$} $(\ell=2\sim 29)$ \\
  \hline 
  Base-$\Lambda$CDM best fit & -- & $1.013$  \\
  \hline
  
  HI & $0$ & $0.971$  \\

 HI $+$ T best fit ($\widetilde{T}_{i} \lesssim 10^{-3}~M_{P}$) & $5.15 \times 10^{-4} $ & $0.765$  \\
 HI $+$ T best fit ($\widetilde{T}_{i} > 10^{-3}~M_{P}$)    & $3.62 \times 10^{-3} $   & $0.762$    \\
  \hline

  HI $+$ T & $1.25 \times 10^{-4} $ & $0.918$  \\

  HI $+$ T & $2.50 \times 10^{-4} $ & $0.829$  \\

  HI $+$ T & $5.00 \times 10^{-4} $ & $0.766$  \\

  HI $+$ T & $1.00 \times 10^{-3} $ & $0.871$  \\

  \hline
  \end{tabular}
  \caption{\label{tab2} $\chi^2$ values for different models using the Planck 2018 data \cite{Planck}. The HI and HI $+$ T denote the standard Higgs inflation and the Higgs inflation starting with a finite temperature, respectively.}\label{chitable}
  \end{table} 

In our model, we can attribute the suppression of $\mathcal{D}^{TT}_{\ell}$ and $P_{\mathcal{R}}(k)$ at large scales to the entropy density term in the denominator of eq.~(\ref{Pk2}). As the temperature increases, the entropy density associated with the Helmholtz free energy will increase as well and significantly change the equation of state of $\chi$. In addition to the temperature effect, eq.~(\ref{Pk4}) shows that $P_{\mathcal{R}}$ is suppressed more when adding up more degrees of freedom in the SM. In practice, we can imagine the whole scenario as follows: if the Universe undergoes the inflation starting with an ultra-high temperature $\widetilde{T}_{i}$ as the horizon crosses the CMB scale $k_{\mathrm{CMB}}^{-1}$, the primordial curvature power spectrum is approximately suppressed by $\widetilde{T}^{-4}$ at the largest scale. Soon after the onset of the inflation, any temperature field $\widetilde{T}$ is immediately redshifted to zero and part of the SM degrees of freedom decouple from the heat bath. The decrease in temperature and the reduction of the active degrees of freedom together diminish the thermal effects and the suppression in the amplitude of power spectrum. At certain point the temperature is so low ({\it i.e.}, $\widetilde{T}< \widetilde{m}$) that the approximation in eq.~(\ref{Pk4}) becomes invalid. Accordingly, we expect that $P_{\mathcal{R}}$ should gradually approach the standard power-law spectrum $P_{\mathcal{R}}(k) \propto k ^{n_{s}-1}$ at small scales.

%==============
\subsection{Can the thermal equilibrium be established?}\label{44}
%==============

Since we have assumed that the Universe is thermalized before inflation, a necessary question to answer is that whether it is possible to reach thermal equilibrium given high initial temperatures in figures \ref{fig:Pk}, \ref{fig:TT} and table \ref{chitable}. In general, particles have to experience sufficient interactions within the time scale that is shorter than the cosmic age. Therefore, the thermal equilibrium can only be established when the interaction rate between particles is larger than the expansion rate of the Universe 
\be \label{thermalize}
\Gamma_{\mathrm{int}} > \widetilde{H}~.
\ee
The interaction rate of the relativistic SM particles with temperature $\widetilde{T}$ above the electroweak scale can be estimated by
\be 
\Gamma_{\mathrm{int}} = n \left\langle \sigma v \right\rangle \approx  \frac{\zeta(3)}{\pi^2} \, g_{*}\widetilde{T}^3 \cdot \frac{\alpha^2}{\widetilde{T}^2} \sim  0.1 g_{*}\alpha^2 \widetilde{T} ~, \label{intcross}
\ee
where $n$, $\sigma$, $v$, $\alpha$ are the number density, cross section, relative velocity and the squared of the coupling strength, respectively. The effective number of relativistic degrees of freedom $g_{*}$ that dominate the interactions with the Higgs field at high energies is about $20$ from $H^0$, $W^{\pm}$, $Z^0$ and $t$, and we adopt $\alpha \sim 10^{-1}$ in the SM  \cite{KolbTurner}. On the other hand, using eqs.~(\ref{Ularge}), (\ref{FE}) and (\ref{Ts4}), we can approximate the expansion rate $\widetilde{H}$ by
\be
\widetilde{H} \approx \frac{1}{\sqrt{3}M_P} \lp U_0 +  \frac{\pi^2}{30}\, \mathcal{G}_{*}  \widetilde{T}^4 \rp^{1/2} ~,  \label{HTU0}
\ee
where the tree-level inflationary plateau $U_0 \sim 10^{-10}\,M_P^4$ and the total number of degrees of freedom of all SM particles is $\mathcal{G}_* \sim 100$. A rough estimation based on eqs.~(\ref{intcross}) and (\ref{HTU0}) and the thermalization criterion in eq.~(\ref{thermalize}) then yields respectively a lower bound and an upper bound on $\widetilde{T}$: 
\be \label{Tthermaleq}
2 \times 10^{-4}~M_P < \widetilde{T}  < 7 \times 10^{-3}~M_P~.  
\ee
Essentially, the two best-fit temperatures we have obtained in table \ref{chitable} favor the consideration of thermal equilibrium.

%=================================================================%
\section{Summary and conclusions} \label{5}
%=================================================================%

The Higgs inflation is by far one of the most successful and theoretically-appealing models of inflation. It is thought that the free energy density of the Higgs field can be non-trivially modified by the finite-temperature effect beyond tree level within the framework of the thermal field theory. Therefore, it is intuitive to think about whether non-zero temperature of the Higgs field, if existing before the inflation, can leave any significant imprint on the CMB anisotropy at large scales.  Through this work, we tackle the problem of thermal effects on the large-scale curvature perturbation for the Higgs inflation.

By assuming that the Higgs field is immersed in a heat bath with a non-zero temperature $\widetilde{T}$ at the outset of inflation, we present the first calculation of the finite-temperature effective potential at the one-loop level based on the inflationary potential $U_{0}(\chi)$ in the Einstein frame. We compute the quantum corrections of the non-minimally coupled field in the Einstein frame for two reasons. First of all, we can bypass the difficulties in including graviton loops to the effective potential. Secondly, the cosmological observables of the Higgs inflation are typically derived in the Einstein frame.  It is more consistent to keep track of all physics in the same conformal frame. Our analyses show that the Coleman-Weinberg effective potential and the quantum instability from the negative mass-squared in the Einstein frame are not important in the inflationary era.  Nevertheless, the thermal effective potential can significantly modify the dynamics and the equation of state of the effective scalar field $\chi$ when more degrees of freedom in the Standard Model are included.

Given the fact that the temperature decreases precipitously throughout the inflationary era, the thermal corrections to $\chi$ are expected to be more influential in the inflaton perturbations at large scales. We find that, owing to the tremendous entropy density originating from the thermal corrections, the primordial curvature power spectrum $P_{\mathcal{R}}(k)$ is approximately enhanced by $\widetilde{T}^4$ at extremely high temperatures $\widetilde{T} \gg 10^{-3} ~M_{P}$ and suppressed by $\widetilde{T}^{-4}$ at the ultra-high temperatures $\widetilde{T} \lesssim 10^{-3} ~M_{P}$. Consequently, $P_{\mathcal{R}}(k)$ would decrease first and then increase if we keep raising the initial temperature $\widetilde{T}_{i}$. Generally speaking, the amplitude of $P_{\mathcal{R}}(k)$ is smaller than the standard power-law prediction for reasonable initial temperatures. We finally calculate the $TT$ angular power spectrum $\mathcal{D}_{\ell}^{TT}$ based on $P_{\mathcal{R}}(k)$ and show that $\mathcal{D}_{\ell}^{TT}$ of the finite-temperature Higgs inflation is also predominantly suppressed at large scales. Using the Planck $2018$ data, the best-fit initial temperature at CMB horizon exit is found to be $ 5.15 \times 10^{-4}~M_{P}$ or $3.62 \times 10^{-3}~M_{P}$. Since the suppression agrees with the deficit of power in the CMB quadrupole moment, our model offers better fit comparing to the best-fit base-$\Lambda$CDM cosmology model. For theoretical consistency, we have also checked that it is possible for the Universe to establish thermal equilibrium with our best-fit initial temperatures before inflation. In conclusion, we have investigated the thermal loop effects on the Higgs inflation at one-loop level based on the finite-temperature field theory and shown that it can serve as a simple scenario to explain the unresolved quadrupole anomaly in CMB.

\acknowledgments

We would like to thank Misao Sasaki and Eibun Senaha for some discussions at the early stage of this project. PWC would like to extend his gratitude to Alan Guth, Yuhsin Tsai and Tommi Tenkanen for useful comments, and Chenxiao Zeng and Chia-Jung Hsu for the helpful discussions about computational tools. CWC would like to thank the New High Energy Theory Center of Rutgers University for their hospitality during his visit, when part of this work was done. This work was supported in part by the Ministry of Science and Technology (MoST) of Taiwan under
grant numbers 104-2628-M-002-014-MY4, 108-2112-M-002-005-MY3 (CWC) and 107-2119-M-001-030 (KWN).

%\paragraph{Note added.} This is also a good position for notes added
%after the paper has been written.

\appendix
\section{Relations of physical quantities between Jordan and Einstein frames}\label{A}

Here we provide some useful relations of the physical quantities and field equations in the JF and EF frames related by the conformal transformation 
\be \label{Appg}
g_{\mu\nu} ~~~ \rightarrow ~~~ \widetilde{g}_{\mu\nu} = f(h) g_{\mu\nu}~. 
\ee 
The ``dot'' and ``prime'' in this section denote the derivatives with respect to the time coordinate $t$ and the non-minimally coupled Higgs field $h$, respectively.

\subsection{Metric and coordinates}
The EF proper time $\tau$ and the EF scale factor $\tilde{a}$ are defined by
  \ba
  ds^2 &&= - dt^2 + a(t)^2 d\vec{x}^2 \nn
  \\
  \rightarrow ~~~ d\tilde{s}^2 &&= - f(h)dt^2 + f(h)a(t)^2 d\vec{x}^2  \equiv - d\tau^2 + \tilde{a}(\tau)^2 d\vec{x}^2~.  \label{Appds}
  \ea
With eq.~(\ref{Appds}), we can further define the EF Hubble expansion rate $\widetilde{H}$ 
  \be \label{AppH}
   H =\frac{\dot{a}}{a} ~~~ \rightarrow ~~~  \widetilde{H} \equiv \frac{1}{\tilde{a}}\lp\frac{d\tilde{a}}{d\tau}\rp = f^{-1/2}H+\frac{1}{2}f^{-3/2}f'\dot{h}~.
  \ee
  
\subsection{Effective quantities}
Under conformal transformation (\ref{Appg}), the effective scalar field $\chi$ with the canonical action can be defined by the non-minimally coupled field $h$ and $f(h)$ through
  \be \label{Apph}
   h ~~~ \rightarrow ~~~  \chi~,~~~\mathrm{where}~~ d\chi=\lp \frac{f+3M_{P}^2 \cdot f'^{\,2}/2}{f^2} \rp^{1/2} dh~.
  \ee
The simple correspondence between the tree-level potential $V$ in the JF and the effective potential $U$ in the EF is
  \be \label{AppV}
  V(h) ~~~ \rightarrow ~~~ U(\chi) \equiv f^{-2}V~.
  \ee

\subsection{Minimally-coupled fields and mass parameters}
To obtain the transformation rules for minimally-coupled scalar field $\varphi$, vector field $A_{\lambda}$, spinor field $\psi_{f}$ and their associated mass parameters, recall the JF action
\be \label{CFJ0}
S_{J}= 
\int d^4x \sqrt{-g} ~\lb \frac{M_{P}^2}{2}f(h)R -\frac{1}{2}g^{\,\mu\nu}\partial_{\mu}h\,\partial_{\nu}h -V(h) + \mathcal{L}_{\mathrm{SM}} \lp g_{\mu\nu},\,\varphi,\,A_{\lambda},\,\psi_{f}\rp   \rb~. 
\ee
We can recast it as the EF action as follows
\be\label{CFE0}
S_{E}= 
\int d^4x \sqrt{-\widetilde{g}} ~\lb \frac{M_{P}^2}{2}\widetilde{R} -\frac{1}{2} \widetilde{g}^{\,\mu\nu}\partial_{\mu}\chi\,\partial_{\nu}\chi - U(\chi) + f^{-2} \mathcal{L}_{\mathrm{SM}} \lp g_{\mu\nu},\,\varphi,\,A_{\lambda},\,\psi_{f}\rp  \rb~,
\ee
where $\widetilde{g}$, $\widetilde{R}$, $\chi$ and $U(\chi)$ are defined by eqs.~(\ref{Appg}), (\ref{Apph}) and (\ref{AppV}), and the SM Lagrangian $\mathcal{L}_{\mathrm{SM}}$ can be expressed as
\ba
&& \mathcal{L}_{\mathrm{SM}} \lp g_{\mu\nu},\,\varphi,\,A_{\lambda},\,\psi_{f}\rp = \mathcal{L}_{1}\lp g_{\mu\nu},\,\varphi \rp + \mathcal{L}_{2}\lp g_{\mu\nu},\,A_{\lambda} \rp + \mathcal{L}_{3}\lp g_{\mu\nu},\,A_{\lambda},\,\psi_{f} \rp ~,  \nn
\\
&& \qquad \bullet ~~~ \mathcal{L}_{1}\lp g_{\mu\nu},\,\varphi \rp =  -\frac{1}{2}g^{\,\mu\nu}\partial_{\mu}\varphi\,\partial_{\nu}\varphi - \frac{1}{2} m_{\varphi}^2 \varphi^2    \nn
\\
&& \qquad \bullet ~~~  \mathcal{L}_{2}\lp g_{\mu\nu},\,A_{\lambda} \rp =  -\frac{1}{4} g^{\alpha\mu} g^{\beta \nu} F_{\mu\nu}F_{\alpha\beta}    \nn
\\
&& \qquad \bullet ~~~  \mathcal{L}_{3}\lp g_{\mu\nu},\,A_{\lambda},\,\psi_{f} \rp =  -i \bar{\psi_{f}}\gamma^{\lambda}\lp \overleftrightarrow{D}_{\lambda}-ieA_{\lambda}  \rp\psi_{f} - m_{f}\bar{\psi_{f}}\psi_{f}    ~.\nn
\ea
Since $S_{J}$ and $S_{E}$ are equivalent, it is straightforward to find the transformation rules for each field and its associated mass parameters by requiring
\be
 \mathcal{L}_{\mathrm{SM}} \lp \widetilde{g}_{\mu\nu},\,\widetilde{\varphi},\,\widetilde{A}_{\lambda},\,\widetilde{\psi}_{f}\rp = f^{-2}  \mathcal{L}_{\mathrm{SM}} \lp g_{\mu\nu},\,\varphi,\,A_{\lambda},\,\psi_{f}\rp   \nn
\ee
and keeping the kinetic terms of each field to its canonical form in $4$-dimensional spacetime:
\ba 
\lp \varphi,\,m_{\varphi} \rp ~~~&&\rightarrow~~~ \lp \widetilde{\varphi} = f^{-1/2} \,\varphi ~,~~\widetilde{m}_{\varphi} = f^{-1/2}\,m_{\varphi} \rp
~, \label{CF1}
\\
\lp A_{\lambda},\,m_{\mathrm{gb}} \rp ~~~&&\rightarrow~~~ \lp \widetilde{A}_{\lambda} = A_{\lambda} ~,~~\widetilde{m}_{\mathrm{gb}} = f^{-1/2}\,m_{\mathrm{gb}} \rp
~, \label{CF2}
\\
\lp \psi_{f},\,m_{f} \rp ~~~&&\rightarrow~~~ \lp \widetilde{\psi}_{f} = f^{-3/4}\,\psi_{f}~,~~\widetilde{m}_{f} = f^{-1/2}\,m_{f} \rp
~, \label{CF3}
\\
\gamma^{\lambda} ~~~ &&\rightarrow~~~ \widetilde{\gamma}^{\lambda} = f^{-1/2}\,\gamma^{\lambda} ~. \label{CF4}
\ea
Hence, we have
\ba 
m_{W}^2 &&= \frac{1}{4}g^2 h^2  ~~~~\,\quad \qquad \rightarrow~~~~ \widetilde{m}_{W}^2= f^{-1} \cdot m_{W}^2  ~, \nn
\\
m_{Z}^2&&= \frac{1}{4}(g^2 + g'^2) h^2 ~~~~ \rightarrow~~~~  \widetilde{m}_{Z}^2= f^{-1} \cdot m_{Z}^2   
~, \label{CF5}
\\
m_{\mathrm{quark}}^2 &&= \frac{1}{2}y^2 h^2 ~~~~\,\quad \qquad \rightarrow~~~~ \widetilde{m}_{\mathrm{quark}}^2= f^{-1} \cdot m_{\mathrm{quark}}^2  ~.\nn
\ea

\subsection{Dynamical equations}
Performing the variation of actions (\ref{CFJ0}) and (\ref{CFE0}), we can write down the Friedmann equation and the equation of motion of the fields via eqs.~(\ref{Appds}) and (\ref{AppH}). The Friedmann equations in both frames read
  \ba 
  H^2 &&= -\frac{f'}{f}\dot{h}H+\frac{1}{3M_{P}^2f}\lp\frac{1}{2}\dot{h}^2+V\rp \label{AppFE1}
  \\
  \rightarrow ~~~  \widetilde{H}^2 &&=\frac{1}{3M_{P}^2}\lb \frac{1}{2}\lp\frac{d\chi}{d\tau}\rp^2+U(\chi) \rb~, \label{AppFE2}
  \ea
while the equations of motion for $h$ and $\chi$ are
  \ba
   &&\ddot{h}+3H\dot{h}   =\frac{-1}{1+\frac{3M_{P}^2f'^2}{2f}}\cdot\lb V'+\frac{f'}{2f} \lp 3M_{P}^2f''\dot{h}^2+\dot{h}^2-4V  \rp \rb \label{AppEOM1}
  \\
  \rightarrow ~~~ &&\frac{d^2\chi}{d\tau^2}+3\widetilde{H}\cdot\lp\frac{d\chi}{d\tau} \rp = -\frac{d}{d\chi}U(\chi)  ~.\label{AppEOM2}
  \ea
Some algebra can identify that eqs.~(\ref{AppFE1}) and (\ref{AppEOM1}) are equivalent to eqs.~(\ref{AppFE2}) and (\ref{AppEOM2}), respectively.

\end{document}